\documentclass[twocolumn]{aastex631}

\usepackage{newtxtext,newtxmath}
\usepackage[T1]{fontenc}
\usepackage{ae,aecompl}

\usepackage{graphicx}
\usepackage{hyperref}
\usepackage{bm}
\usepackage{braket}
\usepackage{natbib}
\usepackage{wasysym}

\shorttitle{Continuous Filament Network of the Local Universe}
\shortauthors{Tugay \& Tarnopolski}

\begin{document}

\title{Continuous Filament Network of the Local Universe}

\author{Anatoliy Tugay}
\email{tugay.anatoliy@gmail.com}
\affiliation{Taras Shevchenko National University of Kyiv\\ Glushkova Ave 4, UA-03127, Kyiv, Ukraine}
\author[0000-0003-4666-0154]{Mariusz Tarnopolski}
\email{mariusz.tarnopolski@umk.pl}
\affiliation{Institute of Astronomy, Faculty of Physics, Astronomy and Informatics, Nicolaus Copernicus University\\ul. Grudzi\k adzka 5, PL-87-100 Toru\'n, Poland}

\begin{abstract}

Simulated galaxy distributions are suitable for developing filament detection algorithms. However, samples of observed galaxies, being of limited size, cause difficulties that lead to a discontinuous distribution of filaments. We created a new galaxy filament catalog composed of a continuous cosmic web with no lone filaments. The core of our approach is a ridge filter used within the framework of image analysis. We considered galaxies from the HyperLeda database with redshifts $0.02\leqslant z\leqslant 0.1$, and in the solid angle $120\degr\leqslant {\rm RA}\leqslant 240\degr$, $0\degr\leqslant {\rm DEC}\leqslant 60\degr$. We divided the sample into 16 two-dimensional celestial projections with redshift bin $\Delta z=0.005$, and compared our continuous filament network with a similar recent catalog covering the same region of the sky. We tested our catalog on two application scenarios. First, we compared the distributions of distance to nearest filament of various astrophysical sources (Seyfert galaxies and other active galactic nuclei, radio galaxies, low surface brightness galaxies, and dwarf galaxies), and found that all source types trace the filaments well, with no systematic differences. Next, among the HyperLeda galaxies, we investigated the dependence of $g-r$ color distribution on distance to nearest filament, and confirmed that early type galaxies are located on average further from the filaments than late type ones.

\end{abstract}

\keywords{
Large-scale structure of the universe (902) --- Cosmic web (330) --- Galaxy clusters (584) --- Catalogs (205)}

\section{Introduction}
\label{sect1}

\subsection{Cosmic Web}
\label{sect1.1}
 
The notion of a cellular large-scale structure (LSS; \citealt{lapparent86}) of the Universe has long been recognized as an important component of the astronomical picture of the world \citep{2002PhR...367....1B}. Optical radiation, coming from the stellar content of a galaxy, is most accessible for observations, but a much larger mass is concentrated in dark matter (DM), whose distribution can be detected by galactic flows \citep{2013AJ....146...69C}, gravitational lensing \citep{1993ApJ...404..441K}, X-ray radiation of clusters \citep{2002ApJ...567..716R}, via the Sunyaev-Zeldovich effect \citep{1980ARA&A..18..537S}, and other methods \citep{2014dmaa.book.....S}. The spatial structure of the cosmic web is reflected primarily in needle-like filaments. Clusters of galaxies occupy a small volume, voids contain a small number of galaxies, and walls contain a number of filaments \citep{2009LNP...665..291V,2013MNRAS.429.1286C}. Through the structure of the cosmic web one can determine the properties of DM \citep{stephon22}, the expansion laws of the Universe, or validate the theory of inflation \citep{2006Natur.440.1137S}. Localization of voids \citep{mao17,voids1} can aid cosmological tests \citep{voids2} and the study of the few galaxies that occupy them \citep{voids3}. 

Two principles lie at the heart of the LSS theory: (i) the hypothesis of the Gaussian distribution of initial density perturbations, and (ii) Zeldovich's approximation within which one can describe their evolution \citep{weygaert02}. Most of the methods for describing the LSS have been tested via numerical simulations \citep{libeskind18,ganesha19,Bonnaire20,Banfi21,yikun22}, and only a few have been applied to actual observational surveys \citep{sousbie11a,sousbie11b,tempel14,chen15,duque22}. A detailed comparison of several modern methods of filament detection is given by \citet{libeskind18}.

A common method for detecting cosmic structures in surveys like the Sloan Digital Sky Survey (SDSS; \citealt{york00,2009ApJS..182..543A}) is a topological method called Discrete Persistent Source Extractor (DisPerSE, \citealt{sousbie11a,sousbie11b}), based on the Delaunay tesselation of the discrete sample of points symbolizing galaxies, which allows to extract filaments, walls, and voids. DisPerSE was applied to illustrate the effect of modified theories of gravity on the length, mass, and thickness of filaments \citep{ho18}. A method of approximating the distribution of galaxies with cylinders (the Bisous model; \citealt{stoica10}) was used by \citet{tempel14} and \citet{muru21}. As a result, the galaxies were grouped into small cylinders (with at least three galaxies inside a cylinder) and were combined into a larger filament net. It was confirmed that about 40 per cent of the mass is gathered in the filaments \citep{forero09,jasche10}, while occupying only 8 per cent of the volume. The detection of filaments was also performed by \citet{chen15} and \citet{duque22} in redshift layers with radial thickness of about 20~Mpc. This approach allows to compensate for the effect of peculiar velocities of galaxies in local gravitational fields by solving a simplified two-dimensional problem, mitigate redshift-space distortions, correlate the projections with maps of cosmic microwave background, trace the evolution of the LSS throughout cosmic time etc. One particular disadvantage is insensitivity to filaments oriented close to the line of sight and inaccurate detections regarding filaments spanning a few neighboring redshift layers,  but the choice of the thickness of the redshift layers is expected to balance the biases and errors. \citet{chen15} used a variant of a ridge filter applied directly to data, and obtained a filament net within $0.05 < z <0.7$, whereas \citet{duque22} extended the sample to redshifts up to $z=2.2$ and employed spherical coordinates to bypass projection effects. The resulting nets follow well the local overdensities of the cosmic web, but in both works contain also some lone, very short filaments, not connected to the rest of the network and not obviously associated to any bigger structure, as well as gaps in the overall network.

The ambiguities of different algorithms and observational restrictions urge the development of more careful methods for detecting and connecting filaments. A way to perform general statistical analyses of the LSS is to generate the apparent distribution of galaxies assuming a certain matter density field
 \citep{voitsekhovski18}. 
This will determine the overall statistical characteristics of the filaments, and therefore describe the LSS of the Universe as a whole. It also allows to produce mock data sets with statistical properties matching those of actual observational surveys. Nowadays, only the closest affiliates of the Local Supercluster \citep{kim16, castignani22}, which are associated with the accumulation of Virgo galaxies, are most prominently identified.

\subsection{Color and Distance Distributions}
\label{sect1.2}

Early galaxy color studies were based on the development and implementation of different photometric systems with appropriate photometric stellar standards\footnote{Later, SDSS magnitudes were intended to be on the AB system; \url{https://www.sdss.org/dr14/algorithms/fluxcal/\#SDSStoAB}.} \citep{vaucouleurs61,sandage78a,fukugita95}. Detailed color studies were conducted for early type (elliptical) galaxies \citep{Bernardi03,Bernardi05}, as well as for late type (spiral) galaxies \citep{tully82,peletier98,fraser16}.

Galaxy color distributions are commonly explained by a halo model which links the galaxy color to the mass of the DM halo \citep{skibba09a,hearin13}. Colors are also correlated with luminosities and morphological types of the galaxies \citep{kodama97,kodama98}. Such relations were analyzed by \citet{mobasher86} and \citet{bower92a,bower92b} for nearby galaxies in Virgo and Coma clusters, and by \citet{lange15} for more distant galaxies at $z<0.1$. An important feature of galaxy colors is a bimodal distribution that relates to spiral and elliptical galaxies \citep{strateva01}. It is clearly seen in SDSS \citep{baldry04}, Millenium Galaxy Catalog \citep{driver06}, as well as for distant galaxies at $z<2.5$ \citep{brammer09}. Bimodality was also found in the two-dimensional color-color distribution \citep{park05}. The bimodal character appears in cosmological computer simulations as well, e.g., Illustris \citep{nelson18}, or Evolution and Assembly of GaLaxies and their Environment \citep{trayford15}. Colors are connected with environment, in dependence on the surface brightness and luminosity \citep{blanton05}. The existence of the green valley, though, has been challenged \citep{eales18}, and explained to arise as a consequence of the Malmquist bias in the submillimetre waveband.

Redder and more massive galaxies are located closer to filaments \citep{luber19}. Surplus of massive red elliptical galaxies in SDSS filaments was analyzed by \citet{kuutma17,kuutma20}, and \citet{welker20} presented similar results for the Sydney Australian Astronomical Observatory Multi-object Integral field spectrograph galaxy survey along with determining the angles between the galaxies' spin and filaments. A corresponding study of X-ray clusters' orientation in relation to the filaments was performed by \citet{shevchenko17}.

\subsection{Outline}
\label{sect1.3}

We present the results of applying a ridge filter within the framework of image analysis to build a continuous net of filaments for radial layers of the galaxy distributions up to redshift $z=0.1$ within a $120\degr\times 60\degr$ solid angle. This paper is organized as follows. Section~\ref{sect2} gives a description of the utilized galaxy sample and discusses its completeness. In Sect.~\ref{sect3} the algorithm for filament detection is outlined. In Sect.~\ref{sect4} we test our approach on a simulated data set. In Sect.~\ref{sect6} we provide a comparison of our filaments with similar outcomes by \citet{chen15}. Section~\ref{sect7} is devoted to validate the extracted filament net through applications: a statistical analysis of distances of various astronomical sources to the nearest filament is performed in Sect.~\ref{sect71}, and in Sect.~\ref{sect72} we examine how the $g-r$ color of galaxies changes with distance from nearest filament. Discussion and concluding remarks are gathered in Sect.~\ref{sect8}. We utilize the cosmological parameters $H_0=70\,{\rm km\,s^{-1}\,Mpc^{-1}}$ and $\Omega=0.32$ throughout.

\section{Sample}
\label{sect2}

We selected the SDSS region by restricting the coordinates within $120\degr\leqslant {\rm RA}\leqslant 240\degr$, $0\degr\leqslant {\rm DEC}\leqslant 60\degr$. Data were taken from the HyperLeda database\footnote{\url{http://leda.univ-lyon1.fr/}} of extragalactic objects \citep{makarov14}. We selected radial redshift layers with thickness $\Delta z=0.005$, which corresponds to 20~Mpc. Such thickness was selected in order to perform a direct comparison with the previous results of \citet[][see Sect.~\ref{sect6}]{chen15}.

\begin{figure*}
\centering
\includegraphics[width=\textwidth]{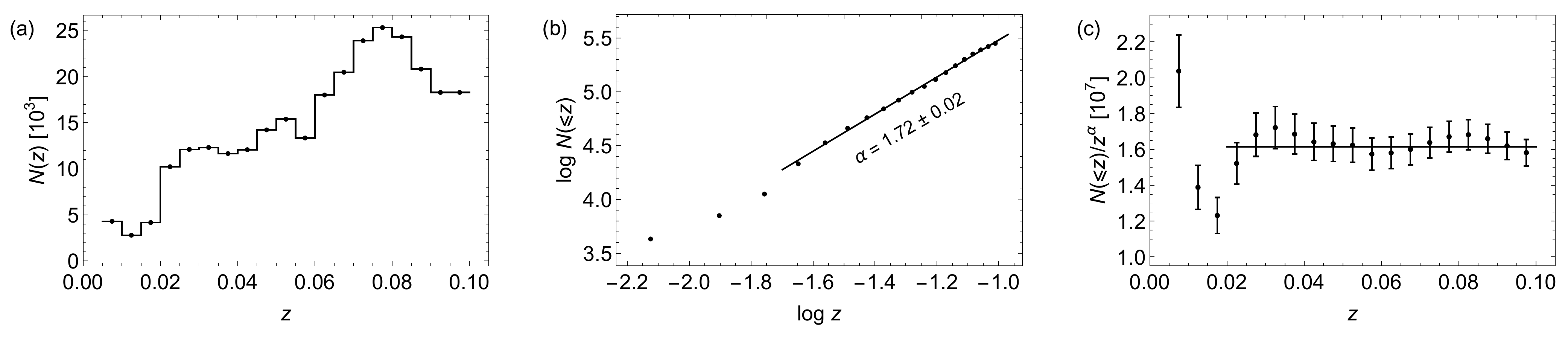}
\caption{(a) Number of galaxies $N(z)$ in each redshift bin. (b) The slope of the linear regression of the number of galaxies $N(\leqslant z)$ up to redshift $z$ in a log-log plot gives the exponent $\alpha$, which then (c) is used to demonstrate the catalog redshift completeness in the range $0.02\leqslant z \leqslant 0.1$. The horizontal line illustrates that $N(\leqslant z)/z^\alpha$ is constant in this range. The error bars in panels (a) and (b), $\Delta N=\sqrt{N}$, are about the size of the graphed points and are taken into account in the calculations. }
\label{n_gal}
\end{figure*}

Extragalactic voids have diameters ranging from 50 to 100~Mpc, so one should find a number of voids spanning a few neighboring layers. We selected all HyperLeda objects with appropriate coordinates and redshifts to get the largest possible data set for filament detection. We found that the number $N(z)$ of galaxies in a redshift layer\footnote{For a redshift bin $z_{\rm low}<z<z_{\rm high}$ we take the midpoint, $(z_{\rm low}+z_{\rm high})/2$, as the {\it representative} redshift value.} starts to decrease slightly at $z\gtrsim 0.08$ (Fig.~\ref{n_gal}(a)). We conclude that regions at $z>0.1$ may not be appropriate for a correct filament detection with the sample at hand, especially given the potential for a growing incompleteness of any data set at larger redshifts. Nevertheless, \citet{chen15} and other authors \citep{sousbie11b,tempel14,2017MNRAS.465.3817M,duque22} presented some results of filament detection at much larger distances.

If galaxies were distributed randomly and uniformly, their cumulative number $N(\leqslant z)$ should increase proportionally to $z^\alpha$ with $\alpha=3$. Due to their clustering in filaments power law index $\alpha$ differs from 3, but should remain constant for $z$ up to some {\it redshift completeness limit}. We therefore construct $N(\leqslant z)$, approximate it by a power law and determine $\alpha$ (Fig.~\ref{n_gal}(b)). Then we examine a ratio $N(\leqslant z)/z^\alpha$, which compares the actual number of galaxies in the sample and the power law approximation (up to the constant of proportionality). We observe that such ratio is constant over a range of redshifts (Fig.~\ref{n_gal}(c)), hence we conclude that the considered catalog is consistent with being complete up to $z=0.1$.

The deviation at $z<0.02$ is due to the domination of local, point-like structures (compared to the cosmic web observed at redshifts $z>0.02$). Point-like structures are single galaxies and clusters and they dominate the most local Universe causing a deviation from the $N(\leqslant z)\propto z^\alpha$ law. The first three radial layers correspond to the Local Supecluster and its boundaries (in particular, many dwarf galaxies in the Local Group at $z<0.01$). We live in the Local Galaxy Supercluster which is a major overdensity of galaxy distribution at $\sim$100 Mpc scales. Our Supercluster is surrounded by thin filaments and voids. The central object of the Local Supercluster, the Virgo galaxy cluster, is located at $z=0.004$, which is below the minimal redshift of our sample. There are 5 known filaments in the Local Supercluster \citep{kim16}, and only one of them falls within our $120\degr\times 60\degr$ field of view. Hence we do not detect significant one-dimensional structures at this scale. The second supercluster, Coma, is separated from us by the Local Void. The Coma cluster, a central object of the corresponding supercluster, lies in the redshift bin $z\in [0.02,0.025]$. At this scale a more uniform cosmic web emerges. In summary, our filament extraction algorithm yielded biased outcomes for $z<0.02$, resulting in a number of apparently spurious detections. Eventually, we analyzed 16 equisized, nonoverlapping redshift bins in the range $0.02\leqslant z\leqslant 0.1$.

The physical extent $\mathcal{D}(\theta, z)$, measured in Mpc, of the structures spanning a given angular size $\theta$ on the sky\footnote{We use $\mathcal{D}$ to denote the sizes or distances spanning the celestial sphere, i.e., located at a given $z$ and perpendicular, in principle, to the line of sight, and denote by $d$ the cosmological distances, i.e., those from Earth to an object located at redshift $z$.} (i.e., a projection onto the celestial sphere) depends on the redshift $z$ as
\begin{equation}
\mathcal{D}(\theta,z) = \frac{\theta\pi}{180}\frac{d_p(z)}{1+z},
\label{eq2}
\end{equation}
where $d_p(z) = \frac{c}{H_0}\int_0^z \frac{dt}{h(t)}$ is the proper distance, with $h(z) = \sqrt{\Omega(1+z)^3+1-\Omega}$, and $\theta$ is measured in degrees. This implies that the scale of detected filaments increases with redshift as well. E.g., the angular length of a $\mathcal{D}=10$~Mpc filament (idealized to be perpendicular to the line of sight) is $\theta = 6.6\degr$ at $z=0.02$, whereas it is $\theta = 1.2\degr$ at $z=0.1$. In other words, the $120\degr\times 60\degr$ solid angle considered corresponds to a $180\,{\rm Mpc}\times 90\,{\rm Mpc}$ patch at $z=0.02$, and a $960\,{\rm Mpc}\times 480\,{\rm Mpc}$ patch at $z=0.1$.


\section{Method}
\label{sect3}

The core of the filament extraction algorithm is the application of a ridge filter \citep{steger98,damon99,lopez15} directly on images of the celestial distribution of galaxies in the Mollweide projection. The functionality of {\sc Mathematica}'s built-in command \texttt{RidgeFilter}\footnote{\url{https://reference.wolfram.com/language/ref/RidgeFilter.html}} is exploited for this purpose. A ridge is, simply put, a long and narrow edge of an elevation (the term corresponding to mountain ridges). In other words, a ridge can be identified if a two-dimensional function has a small second order directional derivative in one direction (along the ridge) and a large derivative in the other direction (perpendicular to the ridge).

The image function $f(x,y)$ at pixel position $(x,y)$ is convolved with a symmetric Gaussian kernel,
\begin{equation}
g(x,y;\sigma) = \frac{1}{2\pi\sigma^2}\exp\left( -\frac{x^2+y^2}{2\sigma^2} \right),
\label{eq3}
\end{equation}
with a set value of $\sigma$ (chosen hereinafter to be $\sigma=30$~pixels, px; see Appendix~\ref{appendix_C}), producing a response $h(x,y)$. The second order structure characterizing ridges can be captured with the Hessian matrix, $\mathbf{H}$, of the function $h$. The ridges are regions with major eigenvalues of $\mathbf{H}$ that are large compared to the minor eigenvalues. In particular, one is interested in computing the main principal curvature at each pixel of the image through the main negative eigenvalue of $\mathbf{H}$ \citep{lindeberg98}.

\begin{figure*}
\centering
\includegraphics[width=\textwidth]{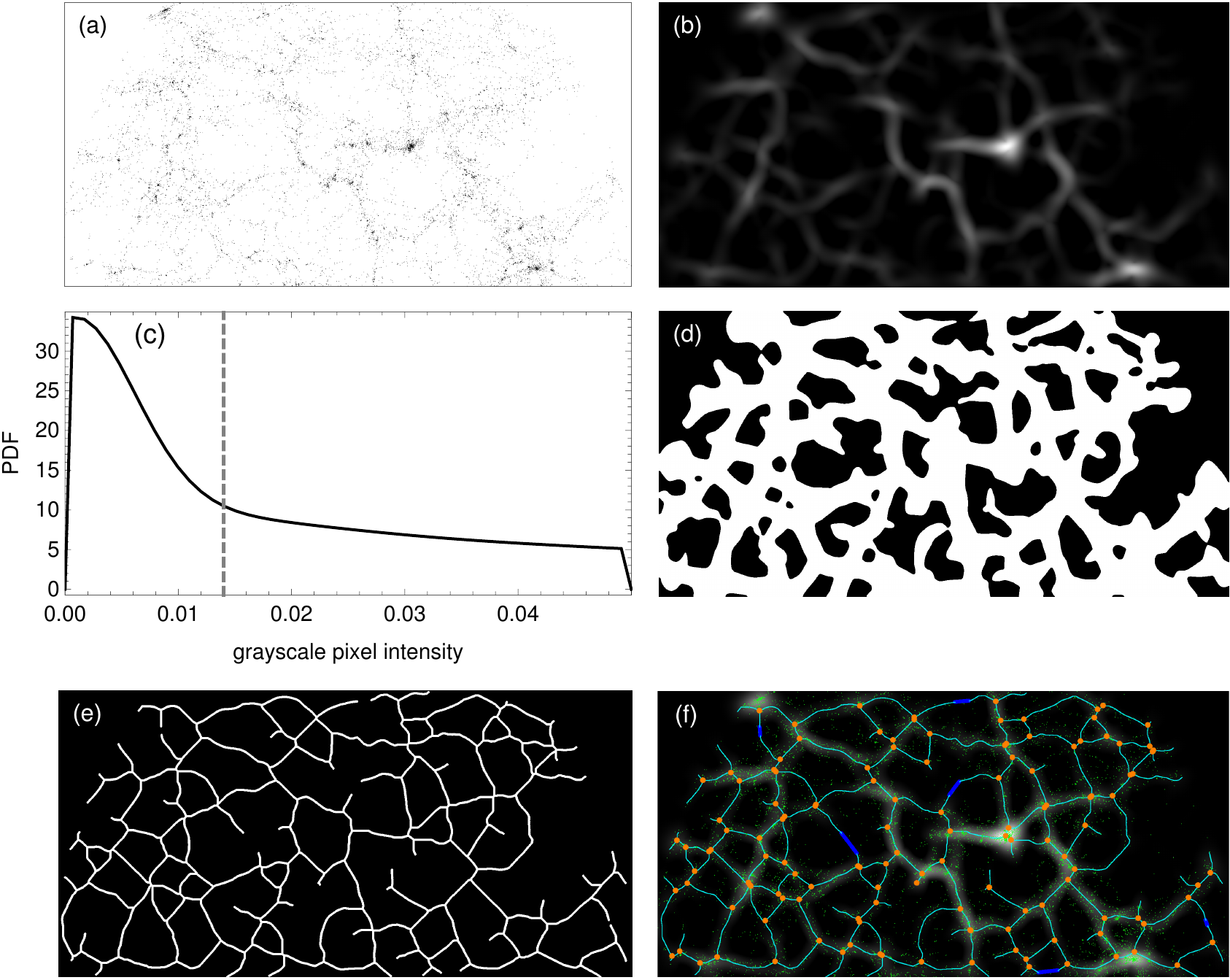}
\caption{ A scheme of the filament extraction algorithm for the redshift bin $z\in[0.025,0.030]$. (a) Celestial distribution of galaxies in the $120\degr\times 60\degr$ field of view. Note that the projection at higher latitudes is appropriately squeezed to take into account the distortion when projecting a spherical sector onto a plane, necessary for the subsequent steps of the image analysis. (b) A direct application of the \texttt{RidgeFilter} yields a grayscale density map, where brighter regions correspond to higher densities. (c) To decide in a binary manner which parts of panel (b) should be associated with filaments, a smooth kernel density of the grayscale pixel intensity is constructed, and the threshold for the separation is chosen as the pixel intensity at which a knee in the probability density function (PDF) is present, marked herein with a vertical dashed line. (d) Pixel intensities exceeding the threshold are colored white (related with overdensities associated with filaments), whereas voids are colored black. (e) A morphological thinning of the binary plot in panel (d) results in the skeleton network of the filamentary LSS. (f) The last step is to connect the tails with other tails via great arcs if they point at each other within a 45 degree cone (blue lines.) Orange dots denote the filament intersections, small green dots mark the location of the galaxies from panel (a), and the grayscale image in the background is the gradient from panel (b).}
\label{cartoon}
\end{figure*}

The crucial steps of the whole procedure are illustrated in Fig.~\ref{cartoon}. The ridge filter is applied to an $1800\,{\rm px}\times 900\,{\rm px}$ image of galaxy locations (Fig.~\ref{cartoon}(a)). The exact dimensions of the image are not that important, since in the next steps it will be blurred and morphologically transformed. In general, the image needs to be of sufficient resolution for the crucial details to be discernible, with $\sigma$ appropriately adjusted. We take into account the distortion of a projection of a spherical sector onto a plane by appropriately {\it squeezing} such Cartesian projection at higher latitudes, therefore sustaining the proper geometry as much as possible. The smaller the considered patch of the sky, the less distorted the geometry.

The \texttt{RidgeFilter} produces a smoothed grayscale image with pixel intensities corresponding to density of the galaxies (color-coded with pixel intensity values equal to 0 for black = low density, and 1 for white = high density; Fig.~\ref{cartoon}(b)). To choose the threshold for the pixel intensity to decide which areas of this grayscale image harbour filaments (i.e., which have densities high enough to be associated with a filament), the smooth kernel distribution of the pixel intensities of the whole image is constructed with the Silverman's rule \citep[][Fig.~\ref{cartoon}(c)]{silverman86}, and the threshold is defined to be the value where the distribution has a knee point, i.e., its curvature is maximal \citep{satopaa11}, marked with the vertical dashed line in Fig.~\ref{cartoon}(c). Values above this threshold are around the filaments, and values below mark the voids. This step yields a binary image with separated regions containing the filamentary surroundings and voids (Fig.~\ref{cartoon}(d), marked as white and black, respectively). Small, isolated filaments (i.e., those with less than $10^4$ px) are deemed artifacts and removed using the \texttt{DeleteSmallComponents}\footnote{\url{https://reference.wolfram.com/language/ref/DeleteSmallComponents.html}} command, and likewise very small voids are closed, i.e., the surrounding filamentary structures are expanded to absorb the void using the \texttt{Closing}\footnote{\url{https://reference.wolfram.com/language/ref/Closing.html}, with \texttt{DiskMatrix[5]} as the kernel.} command. Finally, the filamentary regions are morphologically squeezed (thinned) with the \texttt{Thinning}\footnote{\url{https://reference.wolfram.com/language/ref/Thinning.html}} command to form a net of interconnected one-dimensional filaments (Fig.~\ref{cartoon}(e)).

\begin{figure}
\centering
\includegraphics[width=0.923\columnwidth]{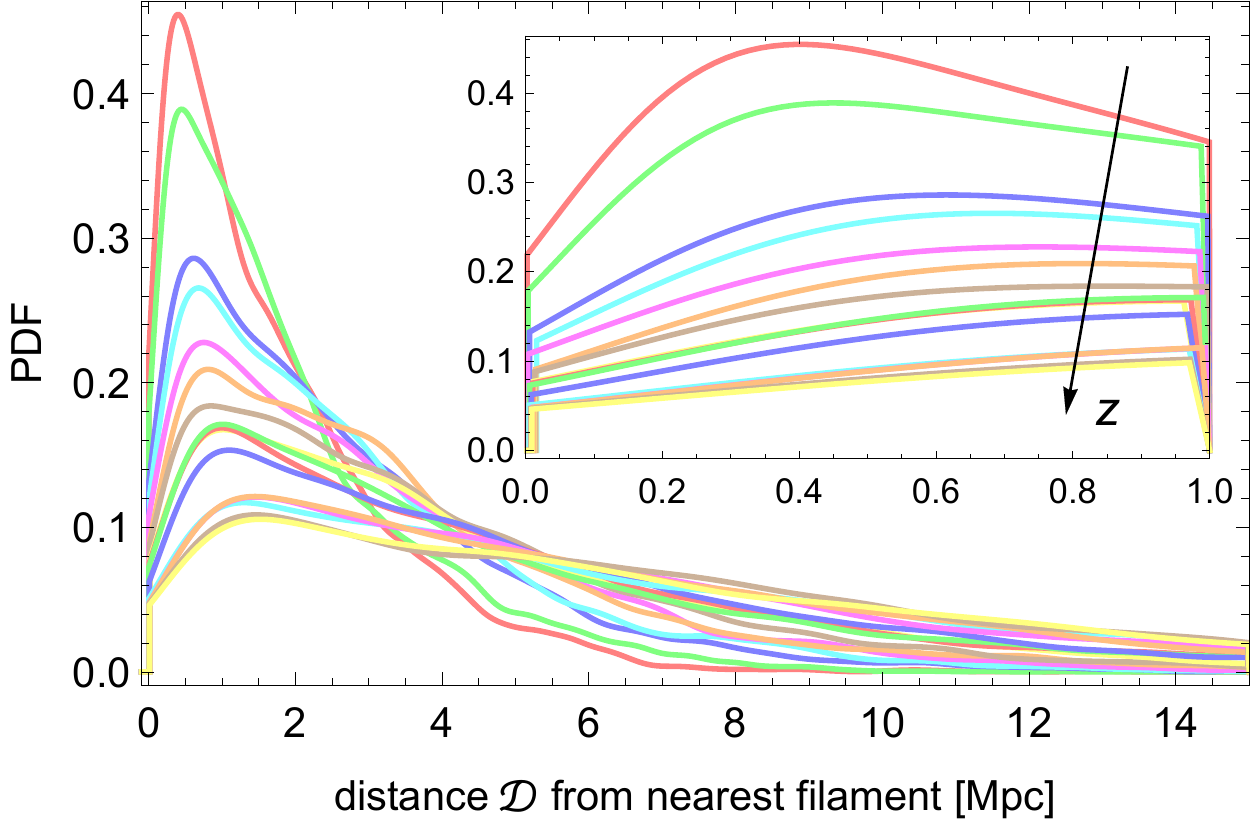}
\caption{Distributions of distance from each HyperLeda galaxy to nearest filament. The inset shows that lower curves consistently correspond to higher redshifts (according to the direction of the arrow). Colors are used only to distinguish between the curves.}
\label{dist_dist}
\end{figure}

At this stage the net still contains some artifacts, like tails, i.e., filaments with one loose end, and pairs of tails pointing at each other but not connected. Tails can be of physical meaning if they are longer than $\mathcal{D}=10$~Mpc -- shorter are removed (with the use of the \texttt{Pruning}\footnote{\url{https://reference.wolfram.com/language/ref/Pruning.html}} command and based on the sizes computed via Eq.~(\ref{eq2})). By physical meaning we understand either tails pointing in the direction of the edge of the field of view, or tails protruding into the voids. The latter can indicate real filaments, which parts might be obscured by the environment or observational constraints. It might be also possible that parts of the filaments have a deficit of bright galaxies, but still contain a significant amount of DM. We therefore keep such tails in the final net. Pairs of tails that point at each other within a 45 degree cone are connected with great arcs. From the resulting final net (Fig.~\ref{cartoon}(f)), the locations of the filaments (cyan and blue lines in Fig.~\ref{cartoon}(f)) and intersections (orange points in Fig.~\ref{cartoon}(f)) are recorded, and constitute our filament catalog (Appendix~\ref{appendix_A}). In Fig.~\ref{dist_dist} the distributions of distance $\mathcal{D}$ to nearest filament are displayed for each redshift bin separately. For higher redshifts the peak of the distribution shifts toward greater distances $\mathcal{D}$, and the distributions become flatter. This illustrates that at higher redshifts the observational biases start to play a role in analyzing the LSS.

We emphasize that despite being based on the ridge formalism via Hessian matrix approach \citep{libeskind18,rost20}, like several other works, a novelty in our overall procedure is to employ the ridge formalism to produce the grayscale images depicting the local galaxy densities, and then proceed exclusively with image analysis techniques that are topological in nature \citep{vandaele20} to distort, transform, cleanse, and eventually extract the backbone of the filament network. Our overall goal was to produce a filament network that is as connected as possible, mitigating the fractured nets commonly obtained by other filament finders. We work on images of the Mollweide projection which is an equal-area map.

The above described procedure was applied to all 16 redshift bins, producing a continuous filament network. The filament catalog is provided as ancillary files (Appendix~\ref{appendix_A}), and a {\sc Mathematica} notebook with the whole implementation (Appendix~\ref{appendix_B}) applied to mock data (see Sect.~\ref{sect4}) is available from \url{https://zenodo.org/record/7971833}. We briefly discuss the potential for a three-dimensional generalization of the algorithm within the image analysis framework in Sect.~\ref{sect8.2}.

\section{Test on mock data}
\label{sect4}

\begin{figure}
\centering
\includegraphics[width=\columnwidth]{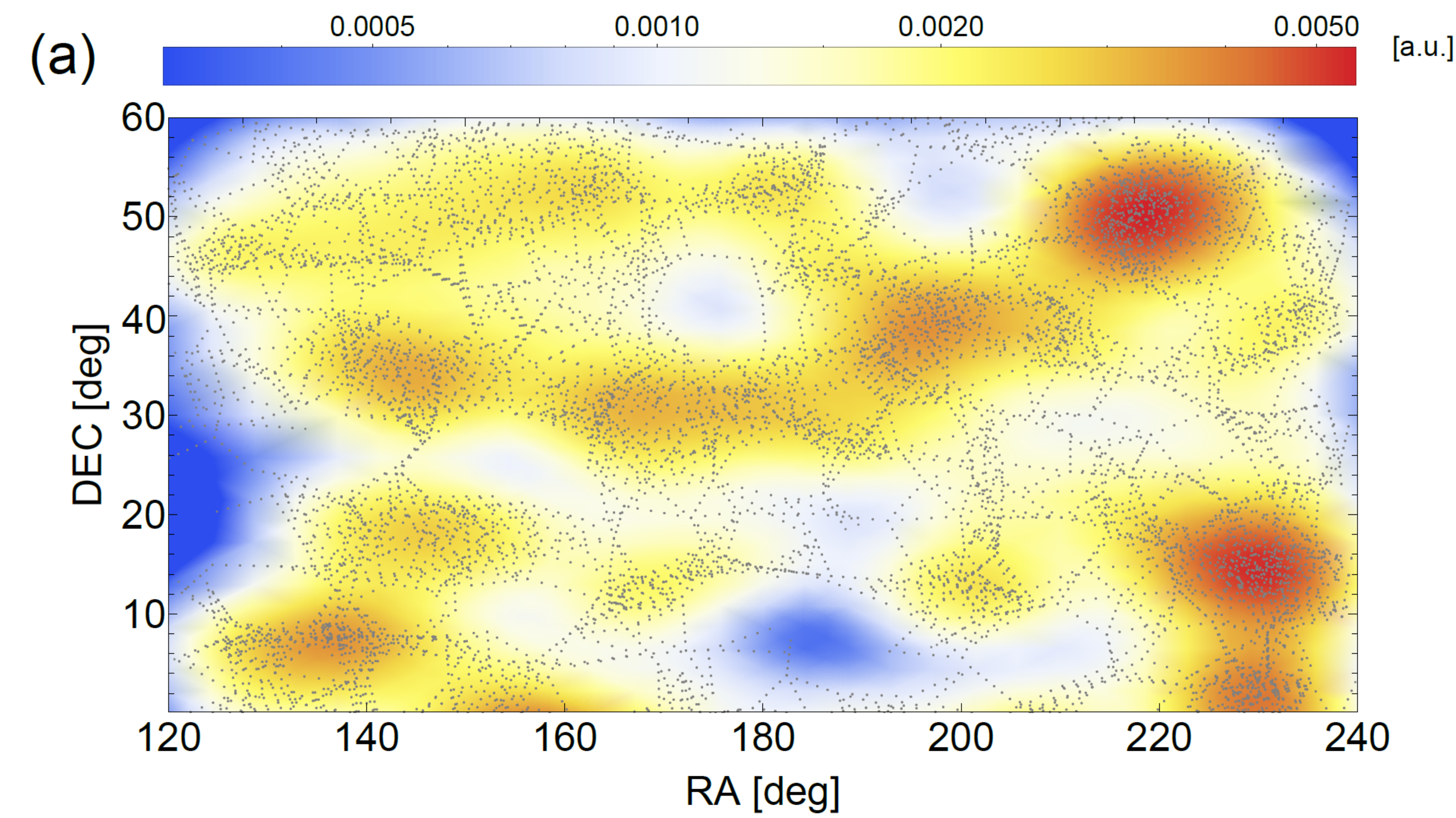}\\
\includegraphics[width=\columnwidth]{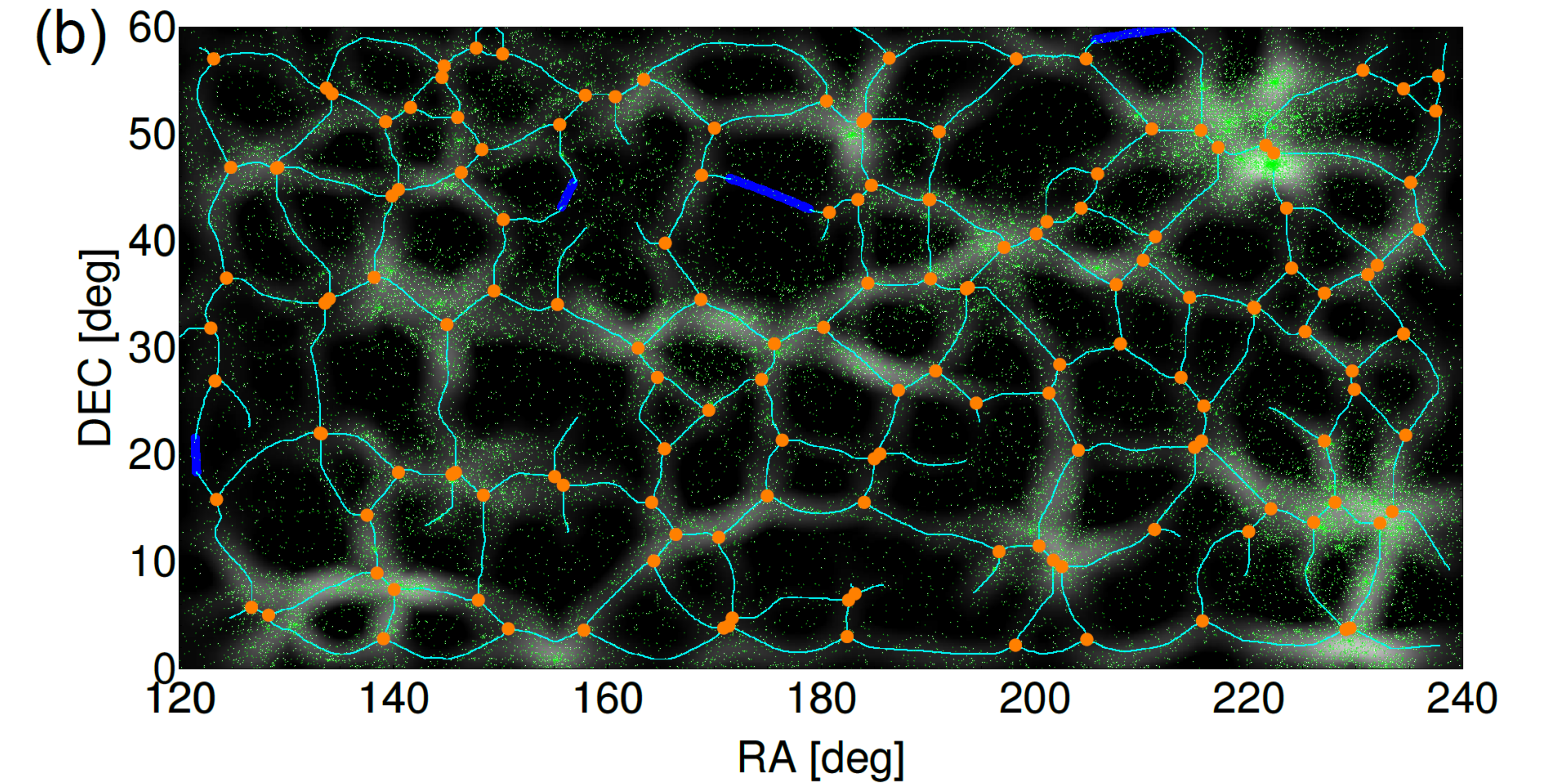}\\
\includegraphics[width=\columnwidth]{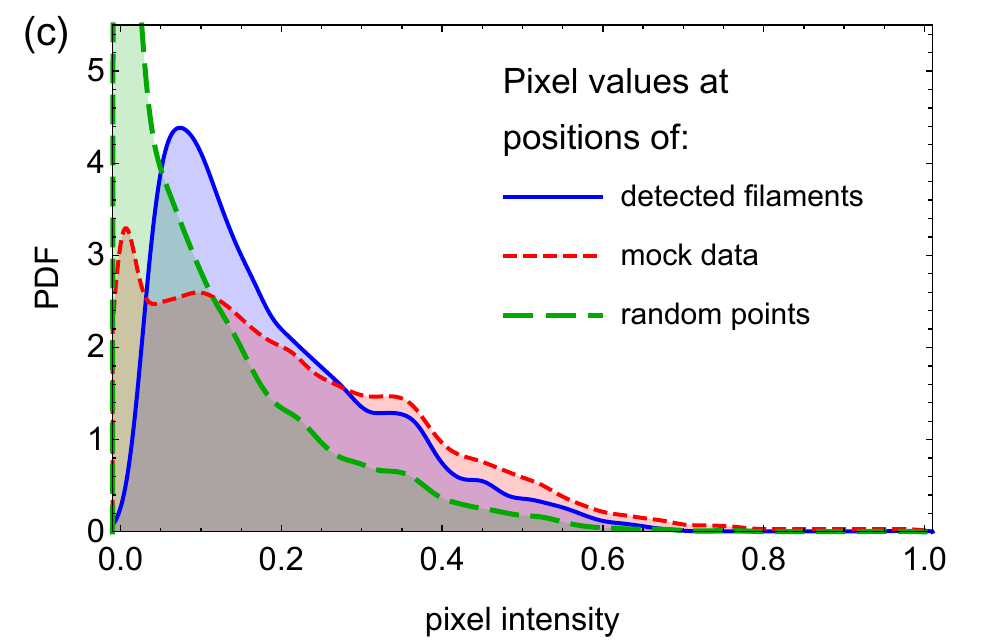}
\caption{The simulated mock LSS. The parameters of random distributions were selected so that the 2pACF corresponds to that of an SDSS layer at a distance of 100~Mpc. (a) The initial points, generated as described in Sect.~\ref{sect4}, overlaid on the corresponding density plot. The (logarithmic) color scale depicts the local concentration of points in arbitrary units (a.u.). (b) The resulting filament network of the final mock data set, obtained with our algorithm described in Sect.~\ref{sect3}. The meaning of colors is the same as in Fig.~\ref{cartoon}(f). Note this plot, and similar subsequent ones, are displayed in Cartesian coordinates for ease of presentation, contrary to the algorithm itself which relies on properly distorted projections (see Sect.~\ref{sect3} and Fig.~\ref{cartoon}). (c) Histograms of the pixel intensities, extracted from the grayscale image of the density plot of the mock data (see the relevant step of the procedure illustrated in Fig.~\ref{cartoon}(b)). The red (short-dashed) line shows the distribution of pixel intensities at locations of the galaxies forming the mock sample. The blue (solid) line is the distribution extracted at locations of the filaments network from panel (b). This distribution is shifted towards values corresponding to higher densities, as it should be expected, since filaments are regions of galaxy overdensities. For comparison, the green (long-dashed) line is formed from the intensities of a random sample of uniformly distributed points.}
\label{mock}
\end{figure}

We tested our algorithm with a random galaxy distribution. We used a sample of randomly distributed galaxies according to \citet{voitsekhovski18}. The sample is generated by randomly distributing clusters (the average distance between them is denoted with $\braket{r}$), connected by filaments if the distance $r$ satisfies $A/2 \braket{r} < r < 2 A \braket{r}$, with $A\sim 1$. The number of galaxies in a cluster is set to be $n = n_{\rm filam}^5/1000$, where $n_{\rm filam}$ is the number of filaments connected to a cluster. The size of a cluster is given by $\theta D$, where $D=1.4\degr$ and $\theta\sim\mathcal{U}(0, \pi/2)$ is drawn from a uniform distribution. A small number of uniformly distributed isolated galaxies was also generated. The precise parameters of distributions were chosen to fulfill the condition of equality of two-point angular correlation function (2pACF) for such generated sample and the real distribution of galaxies from SDSS. The result of the filament extraction algorithm from Sect.~\ref{sect3}, applied to this mock data set, is shown in Fig.~\ref{mock}(a)--(b). Note these plots, and similar subsequent ones, are displayed in Cartesian coordinates for ease of presentation, contrary to the algorithm itself which relies on properly distorted projections (see Sect.~\ref{sect3} and Fig.~\ref{cartoon}). Fig.~\ref{mock}(c) displays in red the distribution of pixel values (in the grayscale color-coded density plot, i.e., the step depicted in Fig.~\ref{cartoon}(b) that illustrates the procedure) at the particular galaxy positions. Shown in blue is the distribution obtained at the locations of the filament network extracted with the algorithm from Sect.~\ref{sect3}. Compared to the pixel intensity distribution of a random sample of $10^4$ points, uniformly distributed on the $120\degr\times 60\degr$ spherical shell (plotted in green), these two distributions are similar in shape, meaning our filaments trace the LSS well, but the distribution of filaments (i.e., blue line) is statistically significantly shifted towards higher intensity values, i.e., corresponds to higher densities of galaxies than the mock data itself (i.e., red line). This is expected since filaments are regions of above average (projected) number density. We therefore conclude that quantitatively the underlying filamentary structure is recovered with a satisfying accuracy.

\section{Comparison with the filament catalog of Chen et al. (2016)}
\label{sect6}

\begin{figure*}
\centering
\includegraphics[width=\textwidth]{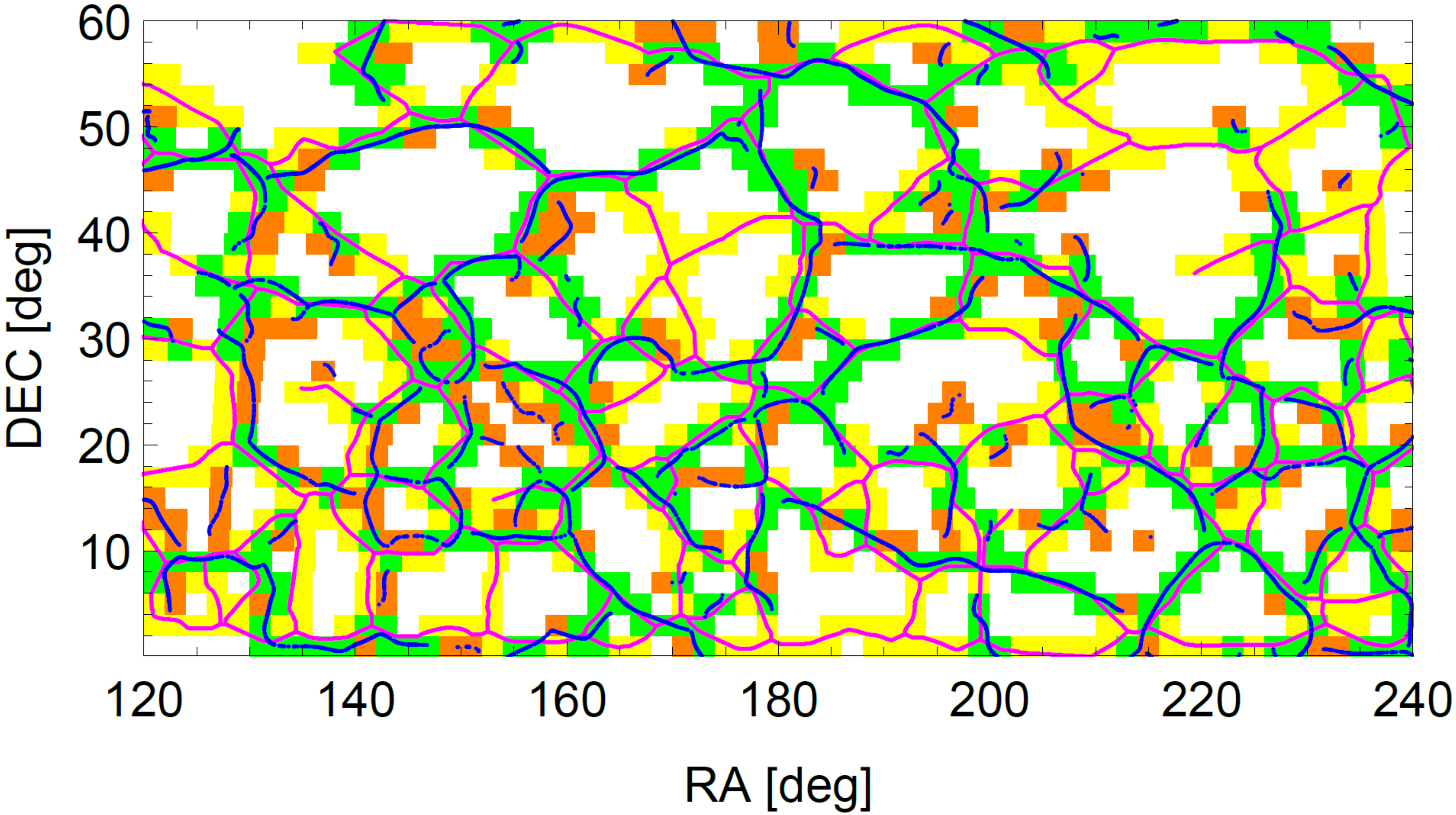}
\caption{Comparison of our filaments (magenta; continuous net) and \citet{chen15} catalog (blue; fragmented net) for the redshift bin $z\in [0.05, 0.055]$. In this case the field is divided into a total of 1486 approximately equal-area boxes, among which 588 do not contain filaments from neither catalog (i.e., are consistently identified as voids). The proportions in Table~\ref{tbl1} refer to the remaining $N_{\rm color}=898$ non-empty boxes, i.e., colored ones. For this example, there are 416 green boxes (containing filaments from both catalogs), 160 orange (only \citealt{chen15} filaments), and 322 yellow ones (only our filaments). See text and Table~\ref{tbl1} for details.}
\label{box_count}
\end{figure*}

In order to compare our filament net with that of \citet{chen15}, we employ the box counting method designed as follows. We divide the $120\degr\times 60\degr$ field of view into boxes of constant step of $2\degr$ in the latitudinal direction, and an appropriate size in the longitudinal one. We choose the latter so that each box has approximately the same angular area, and start with $2\degr\times 2\degr$ boxes at the equatorial strip (i.e. 60 boxes at this first strip), ending with 31 boxes at the last strip. We record the number of cells containing (i) both our and \citet{chen15} filaments, (ii) only \citet{chen15} filaments, and  (iii) only our filaments (color coded with green, orange, and yellow, respectively, in what follows; see Fig.~\ref{box_count}). That is, (i) the same filaments identified by both works, (ii) likely short, detached, spurious filaments, and (iii) places filled in with filaments to maintain continuity of the net. We find that both filament catalogs overlap in about 50\% of the cases (i.e., about 50\% of the boxes are filled by filaments from both catalogs; Table~\ref{tbl1}). Additionally, there are on average 33\% of the boxes that contain only our new filaments, and 18\% with only filaments from \citet{chen15}. This seeming discrepancy comes from different objectives of the two catalogs: we ensured to have the net as connected as possible, with no stray filaments, and very few tails. Therefore, we have filaments in locations where the catalog of \citet{chen15} did not manage to detect, and the catalog of \citet{chen15}, in turn, contains a number of awkwardly lone filamentary scraps, i.e., not connected to the rest of the network, that were discarded from our catalog.

Overall, the $\sim$50\% overlap with the catalog of \citet{chen15} means that we basically detect the same LSS, whereas the fact that 33\% of the boxes contain only filaments from our new catalog, compared to just 18\% containing only those of \citet{chen15} indicates our filaments network covers more densely (or rather more continuously) the given celestial region, which is a result of our main objective that was to construct a filament network as connected as possible, contrary to the one produced by \citet{chen15} and other works (Sect. \ref{sect1.1}).
\begin{table}
\caption{Comparison of our filament catalog with that of \citet{chen15}. The colors denote proportions of cells containing (i) both our and \citet{chen15} filaments (green), (ii) only \citet{chen15} filaments (orange), and (iii) only our filaments (yellow). (See text for details and Fig.~\ref{box_count} for an illustration.)}
\label{tbl1}
\centering
\begin{tabular}{c c c c c}
\hline\hline
redshift bin & $N_{\rm color}$ & green & orange & yellow \\
\hline
$[0.05, 0.055]$ & 898 & 0.46 & 0.18 & 0.36 \\
$[0.055,0.06 ]$ & 863 & 0.50 & 0.21 & 0.29 \\
$[0.06, 0.065]$ & 876 & 0.51 & 0.19 & 0.30 \\
$[0.065,0.07 ]$ & 892 & 0.54 & 0.18 & 0.28 \\
$[0.07, 0.075]$ & 939 & 0.55 & 0.14 & 0.31 \\
$[0.075,0.08 ]$ & 966 & 0.49 & 0.24 & 0.27 \\
$[0.08, 0.085]$ & 905 & 0.40 & 0.15 & 0.45 \\
$[0.085,0.09 ]$ & 901 & 0.46 & 0.15 & 0.39 \\
$[0.09, 0.095]$ & 936 & 0.46 & 0.16 & 0.48 \\
$[0.095,0.1 ]$ & 992 & 0.54 & 0.19 & 0.27 \\
\hline
\end{tabular}
\end{table}

\section{Validation through applications}
\label{sect7}

\subsection{Distance from filaments of various source types}
\label{sect71}

We selected some astrophysical sources to test if different types of objects prefer to reside in different distances $\mathcal{D}$ from the backbone of the filament net. Upon the constraints $120\degr \leqslant {\rm RA} \leqslant 240\degr$, $0\degr \leqslant {\rm DEC} \leqslant 60\degr$, $0.02 \leqslant z \leqslant 0.1$, we gathered the following samples, with respective number of sources fulfilling the above criteria:
\begin{itemize}
\item Seyfert 1 and 2 galaxies \citep{veron} -- 1755 sources. Objects with \texttt{spectrum\_classification=S1} or \texttt{S2} are chosen as secure Seyfert galaxies.
\item AllWISE AGN \citep{secrest15} -- 676 sources from the AllWISE catalog of mid-infrared active galactic nuclei\footnote{\url{https://heasarc.gsfc.nasa.gov/W3Browse/wise/allwiseagn.html}} (AGNs). Only objects with spectroscopically obtained redshifts were used (\texttt{redshift\_flag=s}), which effectively narrowed the range to $0.02 \leqslant z < 0.1$.
\item Radio galaxies \citep{velzen12} -- 67 sources (Fanaroff-Riley class II (FR II) radio galaxies) in the local Universe.
\item FR II -- 148 sources obtained by crossmatching the \citet{velzen15} catalog of FR II radio quasars with SDSS DR7 Main Galaxy \citep{strauss02} and Luminous Red Galaxy \citep{eisenstein01} samples, with a maximum distance between positions of radio source and optical counterpart of 20 arcsec. Redshifts and positions were taken from the SDSS.
\item Low surface brightness galaxies (LSBGs; \citealt{honey18}) -- 146 sources.
LSBGs have typically 5--10 times smaller surface brightness than regular galaxies, and can be of various morphologies (spirals, irregulars, dwarfs). The redshifts were computed from the reported distances\footnote{''Distance (in Mpc) = $V/H_0$, where $V$ is the heliocentric radial velocity in km~s$^{-1}$, for a Hubble constant $H_0=70\,{\rm km\,s^{-1}\,Mpc^{-1}}$'', as stated by \citet{honey18}; for $z\leqslant 0.1$ the difference between $d_p(z)$ and the Hubble law, $d(z)=cz/H_0$, is insignificant (less than 2.5\%), hence we retrieve $z$ via $d_p(z)$ to allow for further distances in future generalizations. } $d$ by solving
\begin{equation}
d = d_p(z)
\label{eq4}
\end{equation}
to place the galaxies in the appropriate redshift bins in our filament catalog.
\item MILLIQUAS AGN \citep{flesch15} -- 9059 sources from the Million Quasars Catalog\footnote{\url{https://heasarc.gsfc.nasa.gov/w3browse/all/milliquas.html}}. Here, like in case of AllWISE AGN, the photometric redshifts (i.e., those rounded to 0.1 according to the catalog description) were discarded, resulting in the effective range $0.02 \leqslant z < 0.1$. The types of sources are: unclassified AGNs, type-1 quasars, BL Lac objects, and narrow emission-line galaxies. 
\item Dwarf galaxies \citep{reines13} -- 88 sources; the stellar mass is within \mbox{$8.6 \leqslant \log \left(M_\star/M_\odot\right) \leqslant 9.48$}. These are all nearby galaxies, with $z\leqslant 0.054$.
\end{itemize}

For comparison, samples of randomly located points (8000 in total, i.e., 500 per redshift bin), drawn from a uniform distribution on the sphere, were generated according to
\begin{subequations}
\begin{align}
x &= 360\degr v, \label{eqA} \\
y &= \frac{180\degr}{\pi} \arcsin(2u-1), \label{eqB}
\end{align}
\end{subequations}
where $v\sim\mathcal{U}\left( \frac{1}{3}, \frac{2}{3} \right)$ and $u\sim\mathcal{U}\left( \frac{1}{2}, \frac{1}{2} \left( \sin\frac{\pi}{3}+1 \right) \right)$ ensure $120\degr \leqslant x \leqslant 240\degr$, $0\degr \leqslant y \leqslant 60\degr$. The centroids of the voids in our filament net (978 locations) were calculated, too, as one cannot be further from a filament than in the center of a void. They are available in our catalog as well (Appendix~\ref{appendix_A}).

\begin{figure}
\centering
\includegraphics[width=\columnwidth]{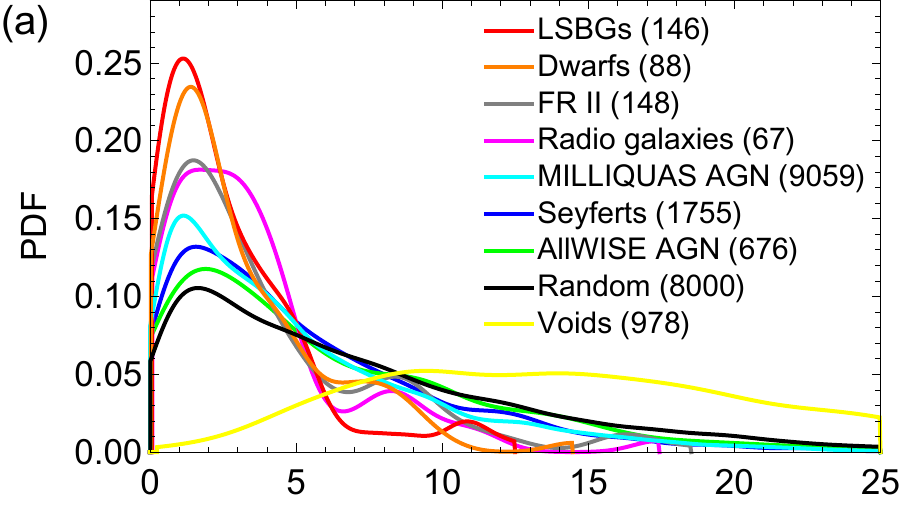}\\
\includegraphics[width=\columnwidth]{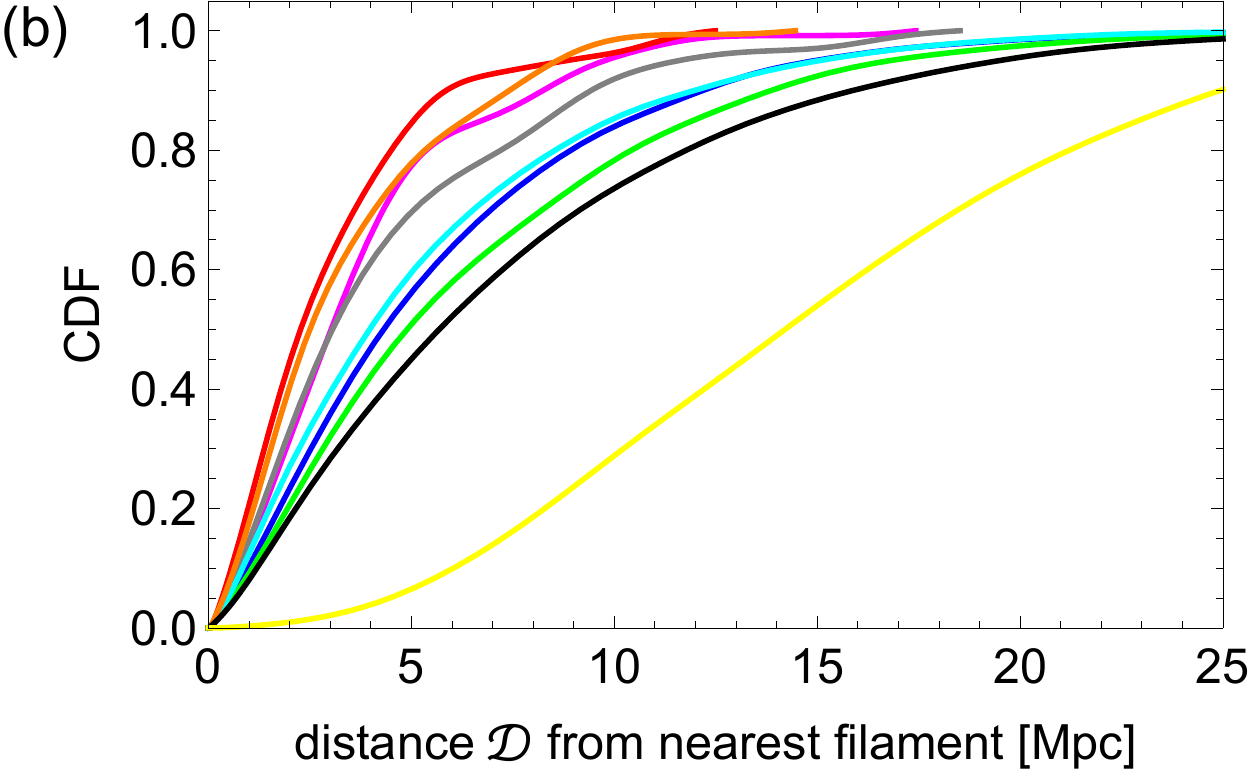}\\
\vspace{0.4cm}\includegraphics[width=\columnwidth]{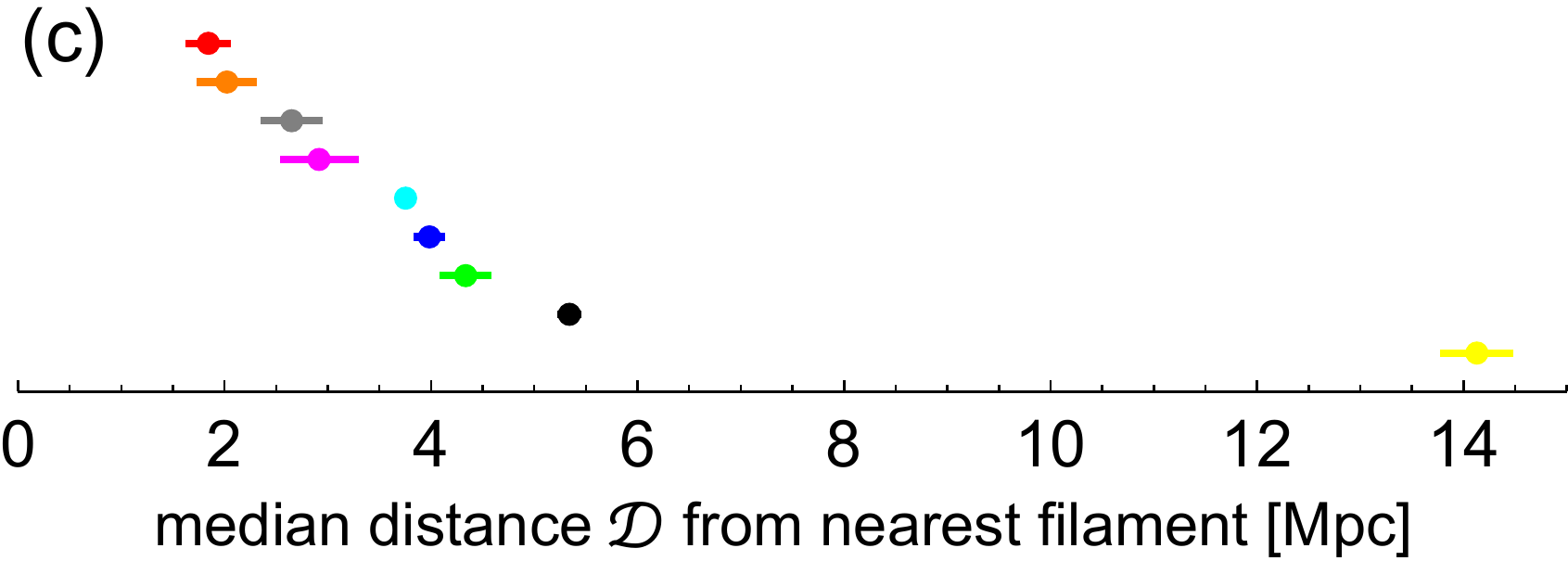}\\
\vspace{0.4cm}\includegraphics[width=\columnwidth]{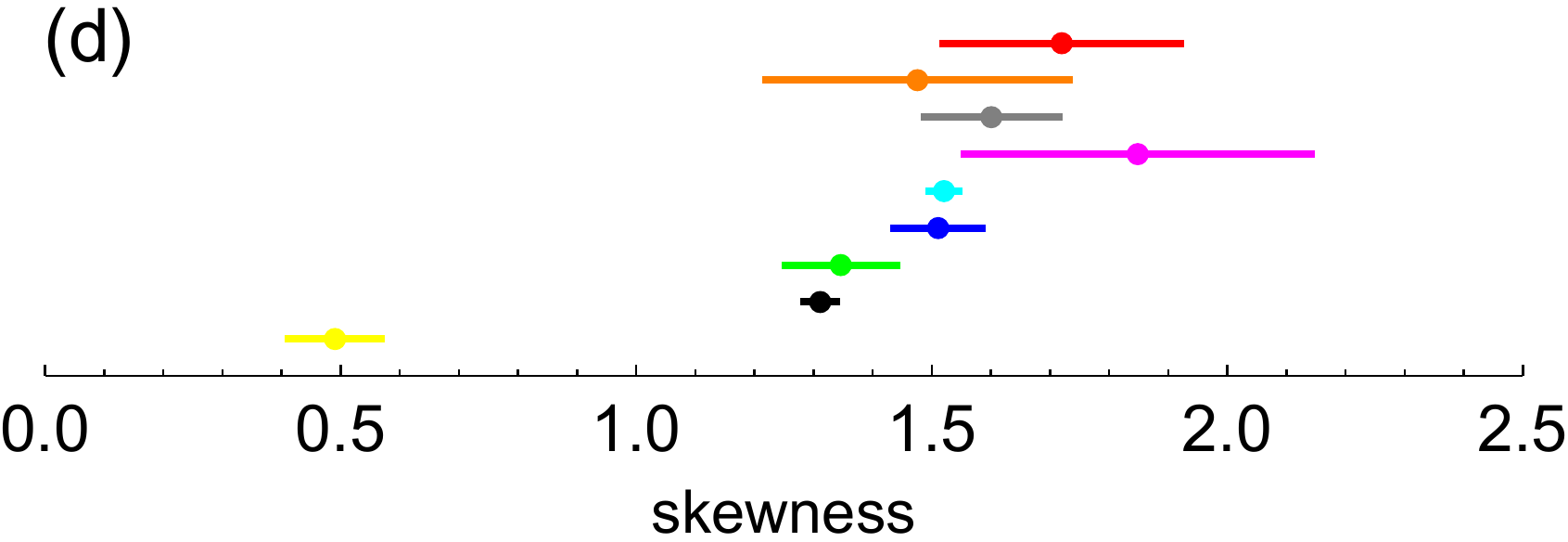}
\caption{Distributions of distance $\mathcal{D}$ to nearest filament for various types of sources: (a) PDFs, (b) cumulative distribution functions (CDFs), (c) medians, and (d) skewnesses. The legend from (a) applies to all panels, and is ordered according to the heights of the PDFs' peaks.}
\label{stats}
\end{figure}

The results are displayed in Fig.~\ref{stats}. All PDFs of the distances $\mathcal{D}$ from the nearest filament exhibit a narrow peak at about $1-2$~Mpc. This is in agreement with the typical widths of simulated filaments \citep{colberg05,gallaraga20}. The PDF of AllWISE AGN seems quite close to the random sample. The medians $m$ and skewnesses $s$ are calculated from the samples, and their sample standard errors (SEs) are
\begin{equation}
SE(m) = \frac{1}{2\sqrt{N}f(m)},
\end{equation}
where $f(m)$ is the PDF evaluated at $m$ \citep{rider60}, and
\begin{equation}
SE(s) = \sqrt{\frac{6N(N-1)}{(N-2)(N+1)(N+3)}}
\end{equation}
for a sample size $N$ \citep{matalas68}.
The skewness measure is important, as for skewed distributions the mean, median, and mode do not align. The mean of a skewed distribution is uninformative in most cases, hence we report the medians instead. For all source types, the medians are significantly smaller than for a random sample (including AllWISE AGN). The PDF for voids is the most symmetric one, with a median of about 14~Mpc.
The Seyfert, AllWISE, and MILLIQUAS samples (i.e., AGNs) have their medians at about 4~Mpc. For the remaining samples, the medians are at $2-3$~Mpc, with LSBGs and dwarfs slightly closer to filaments than FR II radio galaxies. The skewnesses of the AllWISE and random samples are comparable, but the difference in their medians indicate different weight in the tails of the distributions. Overall, one can say that all types of objects examined herein tend to gather near the filaments, and no striking differences between them are visible.

\begin{figure*}
\centering
\includegraphics[width=0.94\textwidth]{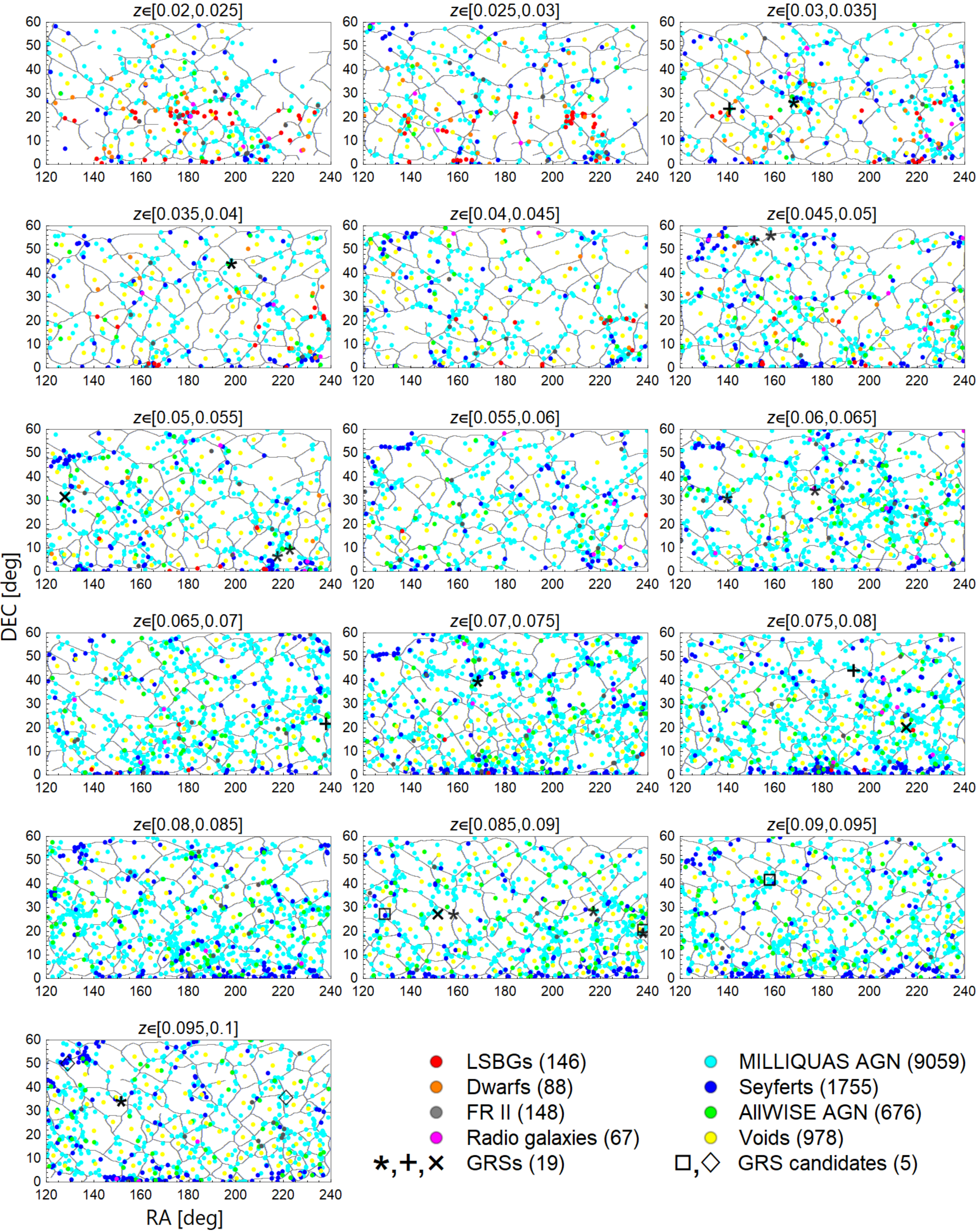}
\caption{Locations of the various source types relative to our filament network in the respective redshift bins. Color codes correspond to Fig.~\ref{stats}. '$\APLstar$' symbols denote GRSs from \citet{kuzmicz18}, '$+$' symbols -- new GRSs from ROGUE~I \citep{koziel20}, '$\square$' symbols -- GRS candidates from ROGUE~I, '$\times$' symbols -- new GRSs from ROGUE~II, '$\Diamond$' symbols -- GRS candidates from ROGUE~II. }
\label{sources}
\end{figure*}

All above sources are plotted on a celestial map in consecutive redshift bins in Fig.~\ref{sources}. We observe that the objects trace our filament net rather tightly. Finally, the catalog of giant radio sources (GRSs, \citealt{kuzmicz18}), while containing 349 sources in total, has only 13 sources fulfilling our RA, DEC, and $z$ constraints.
This sample is too small to make any statistical inferences, but we place the relevant GRSs on the celestial maps for illustration. Likewise, we display the three new GRSs and three GRS candidates from the Radio sources associated with Optical Galaxies and having Unresolved or Extended morphologies I (ROGUE~I) catalog\footnote{\url{http://rogue.oa.uj.edu.pl}} \citep{koziel20}, and three (two) GRSs (GRS candidates) from the upcoming ROGUE~II catalog (N.~\.Zywucka, priv. comm.).

\subsection{Relation between color and distance from filament}
\label{sect72}

We ask the question whether the $g-r$ color of the galaxies changes with distance $\mathcal{D}$ from filaments. For that purpose, we examine the color distributions in five consecutive distance bins: for the first redshift bin, i.e., $z\in[0.02, 0.025]$, we choose galaxies falling in the ranges of $\mathcal{D}=0-1$~Mpc, $1-2$~Mpc, $2-3$~Mpc, $3-4$~Mpc, and $4-5$~Mpc. In subsequent redshift bins, the distance bin sizes are increased by a factor of $16/15$, so that for the last redshift bin, $z\in[0.095, 0.1]$, the distance bins are $\mathcal{D}=0-2$~Mpc, $2-4$~Mpc, $4-6$~Mpc, $6-8$~Mpc, and $8-10$~Mpc. This adaptive binning is justified by the observation that the number of galaxies in each redshift bin is quite similar (Fig.~\ref{n_gal}(a)), we have a constant field of view ($120\degr\times 60\degr$), so that if a size of $\mathcal{D}=1$~Mpc at redshift $z=0.02$ has an angular size $\theta$, a size of $\mathcal{D}=1$~Mpc at $z=0.1$ becomes $\theta/5$. The distance bins are chosen so that a similar number of galaxies is in each, for corresponding redshift bins.

\begin{figure*}
\centering
\includegraphics[width=\textwidth]{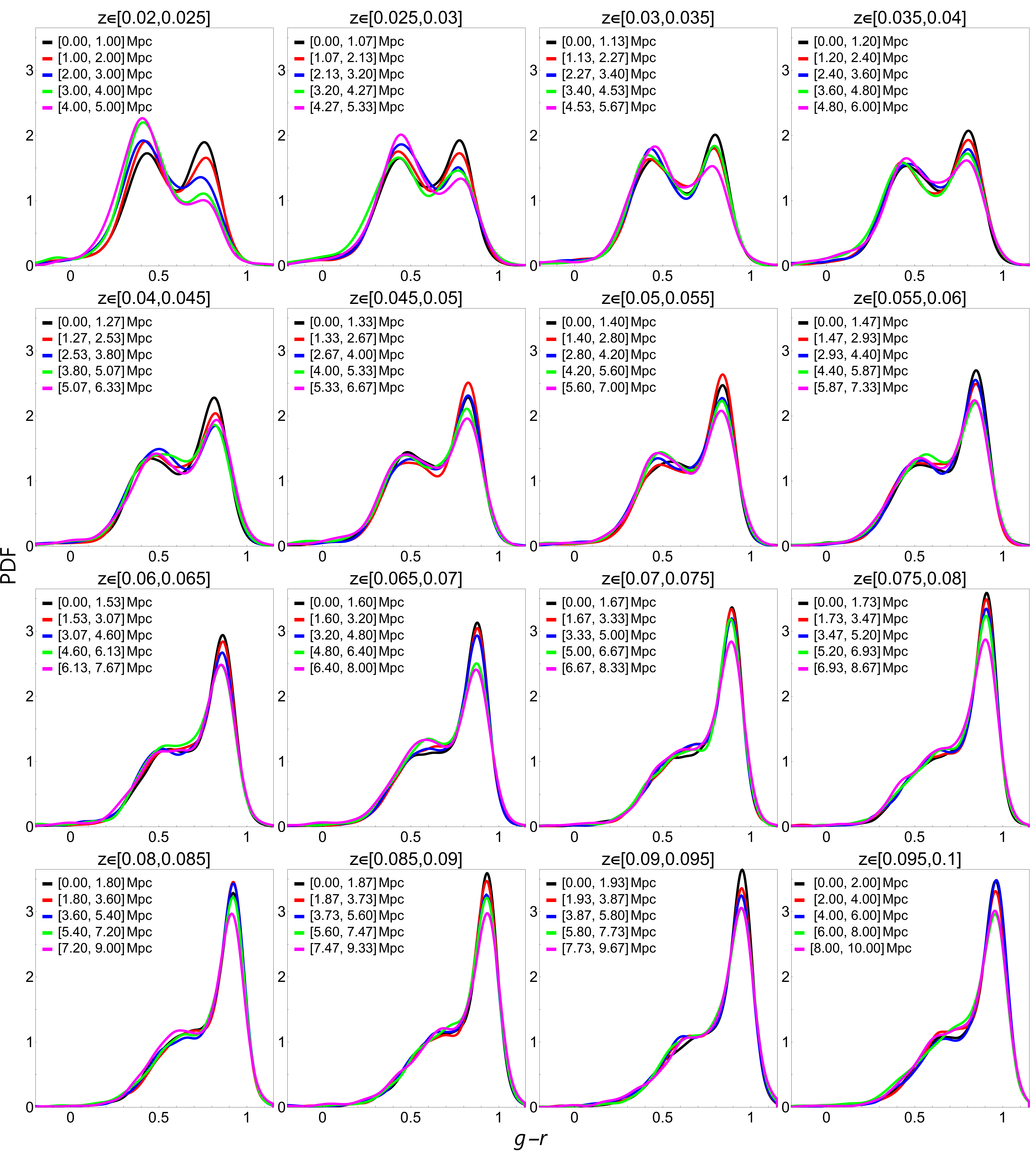}
\caption{Distributions of the $g-r$ color in ascending ranges of distance from nearest filament, for increasing redshift bins.}
\label{colors}
\end{figure*}

Figure~\ref{colors} shows the resulting color distributions. The left peak in each panel corresponds to the blue cloud, while the right peak is the red sequence. We observe the well-known reddening with increasing redshift \citep{sandage73}. It appears that the ratio of blue to red galaxies increases with distance $\mathcal{D}$ to filament. This effect has the following physical explanation \citep{hogg04,berlind05}: galaxies in filaments have more interactions with neighbors, especially at filament intersections and in clusters. Interaction of galaxies with each other \citep{schweizer92,kannappan09}, or even with dense intergalactic medium \citep{mcnamara92}, accelerates the star formation \citep{bower98} and eventually leads to exhaustion of interstellar gas. As a result not much of the interstellar gas remains \citep{kauffmann98}. A blue color indicates the presence of several blue stars within: the brightest, most massive, and short lived stars. A blue color is typically an indicator of current star formation rate \citep{searle73}. Isolated galaxies, e.g., in voids, did not experience much interaction in their past, so they spend their gas on star formation continuously and with low rate. Otherwise, in dense regions of LSS many galaxies lost most of their gas in interactions, so they have a small number of young blue stars, and these galaxies are seen as red. Similar environment dependencies were studied for different samples of SDSS galaxies \citep{ball08}, and in particular in the Galaxy Zoo project \citep{bamford09,skibba09b,darg10}.

\begin{figure*}
\centering
\includegraphics[width=0.495\textwidth]{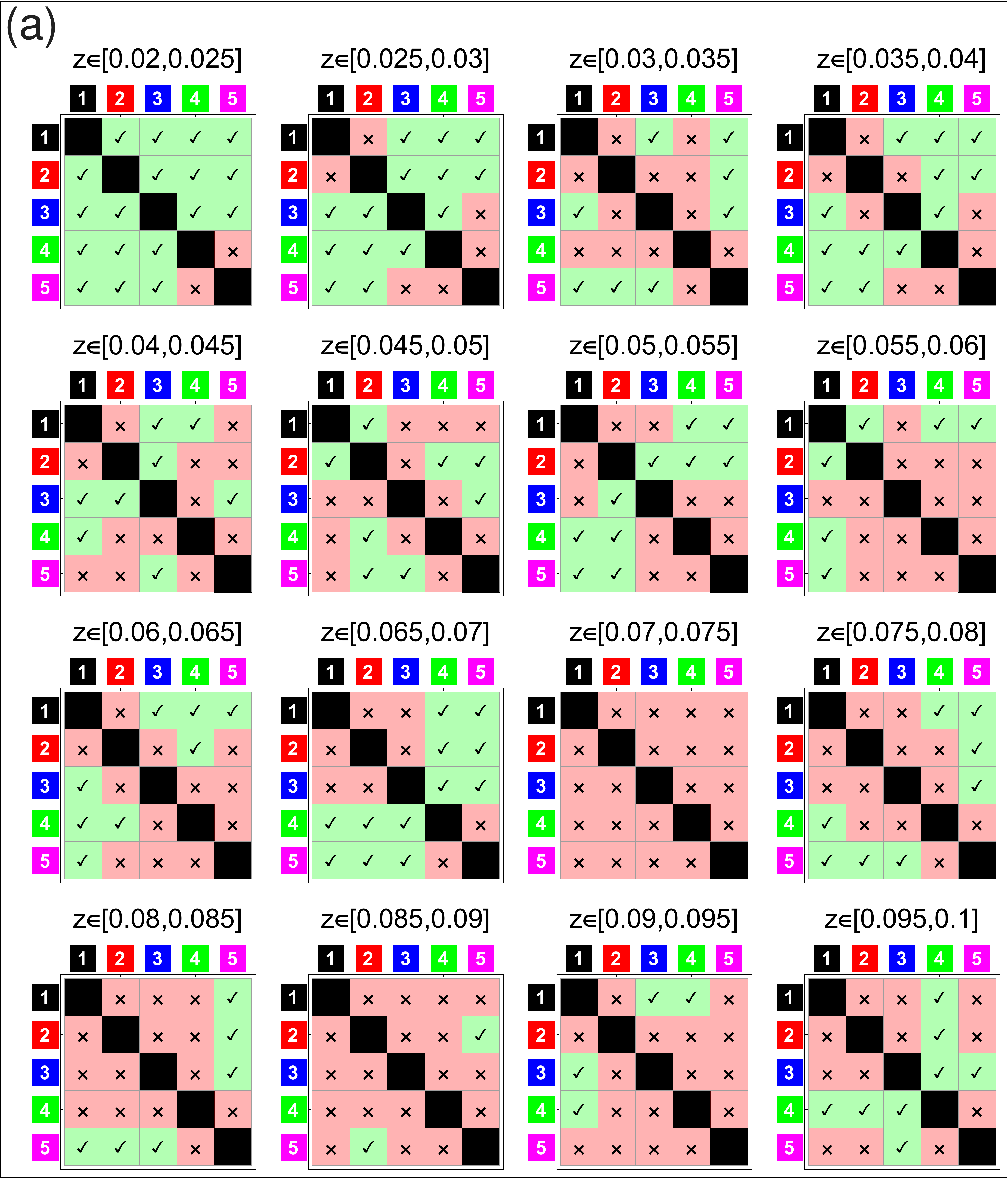}
\includegraphics[width=0.495\textwidth]{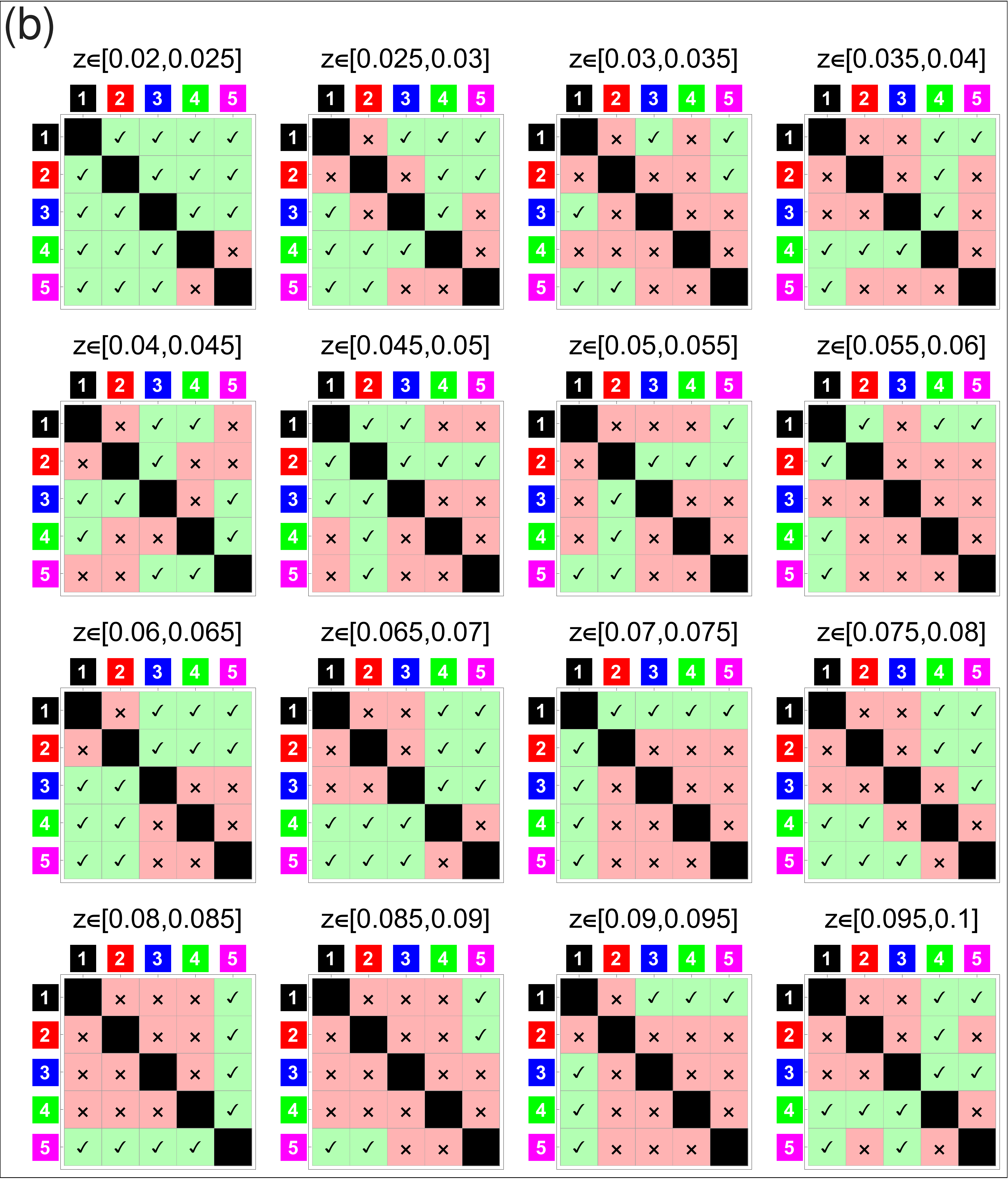}
\caption{ Statistical testing of $g-r$ color distributions. Results of the (a) KS test and the (b) AD test in consecutive redshift bins. In each bin, the labels 1 to 5 denote the subsequent distance ranges from filaments, color coded the same way as in Fig.~\ref{colors}, i.e. 1 -- closest to filaments, 5 -- furthest. The check marks on the light green background denote instances when the two corresponding $g-r$ distributions are statistically different, and the cross marks on the light red background mark cases when the null hypothesis that the considered distributions are the same cannot be rejected. }
\label{KS_AD_tests}
\end{figure*}

To verify whether this effect is statistically significant in our sample, we performed Kolmogorov-Smirnov (KS) and Anderson-Darling (AD) tests in the following way: we perform pairwise tests between all five color distributions in each redshift bin. E.g., in the first bin, $z\in[0.02,0.025]$, we first compare the colors of galaxies closest to the filaments, i.e., those within a distance range of $\mathcal{D}=0-1$~Mpc, with the color distribution of those falling in the second closest range, $\mathcal{D}=1-2$~Mpc, and record the output of the tests (i.e., reject or not reject the null hypothesis). This is then repeated for the first distance range and the third one, second and third, and so on, eventually testing all pairs, and similarly for the subsequent redshift bins. The results are gathered in Fig.~\ref{KS_AD_tests}. The labels numbered 1 to 5 denote the increasing ranges of distance $\mathcal{D}$, color-coded the same way as in Fig.~\ref{colors}. The check marks indicate instances when the tests imply that the respective distributions are indeed different, and cross marks denote cases when the null hypothesis that the distributions are the same cannot be rejected. While the effect of increasing ratio of blue to red galaxies with distance $\mathcal{D}$ to filament is rather clear in the closest redshift bins, it becomes much less prominent for greater redshifts. This is likely associated with insufficient sample size for greater cosmological distances, other selection biases, and observational uncertainties. On the other hand, these outcomes are generally consistent with those of \citet{lee21} who found that the $g-r$ color correlates with distance $\mathcal{D}$ from filaments around the Virgo cluster, and \citet{castignani22} who demonstrated that early-type galaxies are indeed located closer to filaments than late-type galaxies.

\section{Concluding Remarks}
\label{sect8}

\subsection{Outlook}
\label{sect8.1}

Different methods applied to large observational surveys yield different decomposition of the LSS into filaments \citep{rost20}. One can see that segmentation of the same sample of galaxies results in significantly different cosmic web; in particular, with the filament network composed of many disjoint parts, the physical reality of a particular map of the LSS is uncertain. To reliably detect real filaments, new robust methods for their detection should be developed, possibly including other observational strategies. One such approach might be a direct observation of filaments via the 21~cm H\,{\footnotesize I} line, possible with future giant radio telescope arrays like the Square Kilometre Array \citep{kooistra17}. A detailed filament network is an essential tool for modern cosmography, and can be used for a topological description of LSS.

\subsection{Generalization}
\label{sect8.2}

The ridge detection algorithm from Sect.~\ref{sect3} readily generalizes to three-dimensional spaces \citep{lindeberg98}. The framework of image analysis that we built around it can, in principle, be applied to three-dimensional images (i.e., stacks of two-dimensional images), composed of voxels instead of pixels, as well. We leave the implementation and exploration of such a generalization to a future work. It should be noted that the uncertainties in redshift are larger than those of celestial positions, leading to very asymmetric uncertainty distributions in the physical space and posing difficulties in constructing a truly three-dimensional LSS. The benefits of two- or three-dimensional catalogs depend on the particular application and scientific objectives.

\newpage
\subsection{Summary}
\label{sect8.3}

We compiled a catalog of filaments within a solid angle $120\degr\leqslant {\rm RA}\leqslant 240\degr$, $0\degr\leqslant {\rm DEC}\leqslant 60\degr$, covering the redshift range $0.02\leqslant z\leqslant 0.1$ with 16 equisized nonoverlapping bins in celestial projections. These filaments form a continuous net, with no stray filaments, and very few tails. This continuity is an advantage of our LSS model compared to previous works. Our LSS model is simpler in its principles than that of, e.g., \citet{sousbie11a,sousbie11b}: instead of a system of doubtful voids with too complicated shapes and filamentary boundaries, we obtained a cleaner network of smooth filaments. As an illustration of potential applications and validation of such a filament network, we compared the distributions of distance $\mathcal{D}$ to nearest filament of various astrophysical sources (Seyfert galaxies and other AGNs, radio galaxies, LSBGs, and dwarf galaxies; Sect.~\ref{sect71}), and the dependence of $g-r$ color distribution on distance $\mathcal{D}$ to nearest filament (Sect.~\ref{sect72}), obtaining outcomes in tune with other studies. The catalog (Appendix~\ref{appendix_A}) is available online as an ancillary file.

\section*{Acknowledgements}
AT thanks Sergey Shevchenko for assistance with text edition and compilation. MT acknowledges support from the First TEAM grant of the Foundation for Polish Science No. POIR.04.04.00-00-5D21/18-00 and the National Science Center through the Sonata grant No. 2021/43/D/ST9/01153, and thanks Marek Jamrozy, Dorota Kozie\l-Wierzbowska, Natalia \.Zywucka, and Ma\l gorzata Bankowicz for fruitful discussions and help with compiling the data.

We acknowledge the usage of the HyperLeda database (\url{http://leda.univ-lyon1.fr}).

Funding for the SDSS and SDSS-II has been provided by the Alfred P. Sloan Foundation, the Participating Institutions, the National Science Foundation, the U.S. Department of Energy, the National Aeronautics and Space Administration, the Japanese Monbukagakusho, the Max Planck Society, and the Higher Education Funding Council for England. The SDSS Web Site is \url{http://www.sdss.org/}.

The SDSS is managed by the Astrophysical Research Consortium for the Participating Institutions. The Participating Institutions are the American Museum of Natural History, Astrophysical Institute Potsdam, University of Basel, University of Cambridge, Case Western Reserve University, University of Chicago, Drexel University, Fermilab, the Institute for Advanced Study, the Japan Participation Group, Johns Hopkins University, the Joint Institute for Nuclear Astrophysics, the Kavli Institute for Particle Astrophysics and Cosmology, the Korean Scientist Group, the Chinese Academy of Sciences (LAMOST), Los Alamos National Laboratory, the Max-Planck-Institute for Astronomy (MPIA), the Max-Planck-Institute for Astrophysics (MPA), New Mexico State University, Ohio State University, University of Pittsburgh, University of Portsmouth, Princeton University, the United States Naval Observatory, and the University of Washington.

\bibliographystyle{mnras}
\bibliography{bibliography3}

\appendix

\section{Filament catalog description}
\label{appendix_A}

The filament catalog consists of three parts. The first (Table~\ref{tabA}(a)) contains lists of points (285,233), densely sampling the locations of the filaments, in each of the redshift bins, $z_{\rm low}<z<z_{\rm high}$, which are indicated by the lower value $z_{\rm low}$ of the redshift range in a given bin. The second (Table~\ref{tabA}(b); same structure as Table~\ref{tabA}(a)) gathers the coordinates (2,139) of the filaments' intersections in each redshift bin, and the third (Table~\ref{tabA}(c); same structure as Tables~\ref{tabA}(a) and (b)) provides the coordinates (978) of the centroids of the voids.

\begin{table}[h!]
	\centering
	\caption{A sample of our filament catalog. Only first four rows are shown here. The full catalog is available online. }
	\label{tabA}
	\begin{tabular}{ccc}
	    \multicolumn{3}{c}{(a) Filaments.}\\
		\hline
		RA [deg] & DEC [deg] & $z_{\rm low}$ \\
		\hline
        168.5792 & 0.0061 & 0.02 \\
        168.6463 & 0.0061 & 0.02 \\
        168.7133 & 0.0680 & 0.02 \\
        168.7803 & 0.1299 & 0.02 \\
		\hline
	\end{tabular}%
	\begin{tabular}{ccc}
	    \multicolumn{3}{c}{(b) Intersections.}\\
		\hline
		RA [deg] & DEC [deg] & $z_{\rm low}$ \\
		\hline
        187.2397 & 57.9614 & 0.02 \\
        123.0491 & 57.7265 & 0.02 \\
        155.4949 & 56.8700 & 0.02 \\
        203.8066 & 56.7925 & 0.02 \\
		\hline
	\end{tabular}%
	\begin{tabular}{ccc}
	    \multicolumn{3}{c}{(c) Void centroids.}\\
		\hline
		RA [deg] & DEC [deg] & $z_{\rm low}$ \\
		\hline
        171.6858 & 5.1904 & 0.02 \\
        202.1221 & 6.3022 & 0.02 \\
        176.9145 & 4.9114 & 0.02 \\
        151.2692 & 26.0427 & 0.02 \\
		\hline
	\end{tabular}
\end{table}

\section{Implementation of the algorithm}
\label{appendix_B}

A {\sc Mathematica} implementation of the algorithm from Sect.~\ref{sect3}, applied to mock data from Sect.~\ref{sect4}, is available online from \url{https://zenodo.org/record/7971833}.

\section{Impact of the free parameter \texorpdfstring{$\sigma$}{\empty} on the filament net}
\label{appendix_C}

\begin{figure*}
\centering
\includegraphics[width=\textwidth]{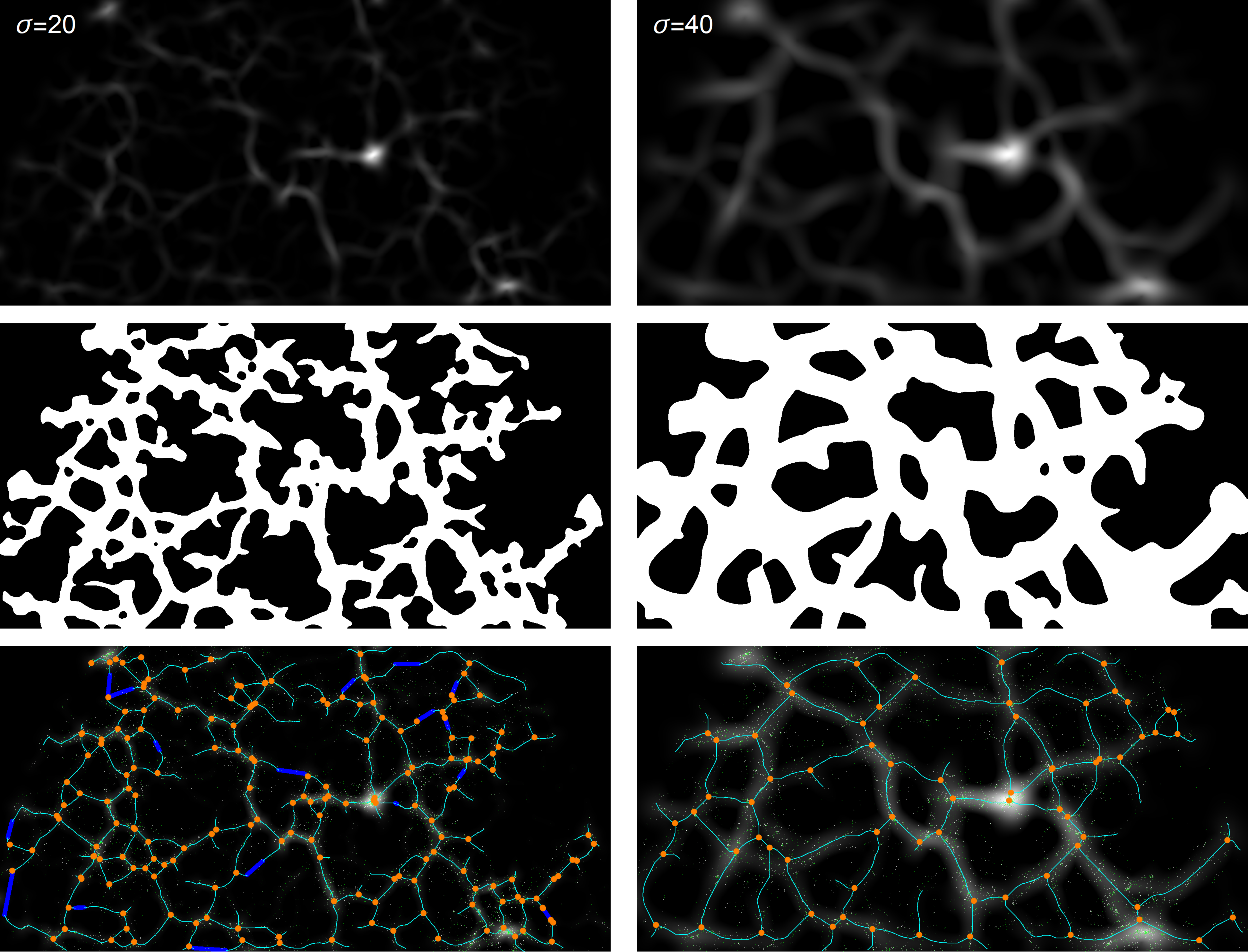}
\caption{ Impact of the value of $\sigma$ from Eq.~(\ref{eq3}) on the resulting filament net. {\it Left}: $\sigma=20$, {\it right}: $\sigma=40$. Rows, from top to bottom, correspond to panels (b), (d), (f), respectively, from Fig.~\ref{cartoon} on which $\sigma=30$ was used. }
\label{cartoon_sigma}
\end{figure*}

Fig.~\ref{cartoon_sigma} shows how the choice of $\sigma$ from Eq.~(\ref{eq3}) influences the backbone of the extracted filament net: smaller $\sigma$ reveal a finer configuration, whilst larger $\sigma$ uncover only the biggest structures. Note that the overall shape, i.e., sizes of the main voids and locations of the longest filaments, is preserved irrespective of $\sigma$.

\end{document}